\documentclass[sigconf]{acmart}

\AtBeginDocument{%
  \providecommand\BibTeX{{%
    \normalfont B\kern-0.5em{\scshape i\kern-0.25em b}\kern-0.8em\TeX}}}

\setcopyright{acmcopyright}
\copyrightyear{2023} 
\acmYear{2023} 
\setcopyright{acmlicensed}\acmConference[WebSci '23]{15th ACM Web Science Conference 2023}{April 30-May 1, 2023}{Evanston, TX, USA}
\acmBooktitle{15th ACM Web Science Conference 2023 (WebSci '23), April 30-May 1, 2023, Evanston, TX, USA}
\acmPrice{15.00}
\acmDOI{10.1145/3578503.3583628}
\acmISBN{979-8-4007-0089-7/23/04}

\setcopyright{none}
\renewcommand\footnotetextcopyrightpermission[1]{}
\setcopyright{none}
\makeatletter
\renewcommand\@formatdoi[1]{\ignorespaces}
\makeatother
\begin{document}

\title{\texttt{$Q_{bias}$} - A Dataset on Media Bias in Search Queries and Query Suggestions}

\author{Fabian Haak}
\email{fabian.haak@th-koeln.de}
\orcid{0000-0002-3392-7860}
\affiliation{%
  \institution{Technische Hochschule Köln}
  \streetaddress{Gustav-Heinemann-Ufer 54}
  \city{Cologne}
  \country{Germany}
  \postcode{50678}
}

\author{Philipp Schaer}
\email{philipp.schaer@th-koeln.de}
\orcid{0000-0002-8817-4632}
\affiliation{%
  \institution{Technische Hochschule Köln}
  \streetaddress{Gustav-Heinemann-Ufer 54}
  \city{Cologne}
  \country{Germany}
  \postcode{50678}
}

\renewcommand{\shortauthors}{Haak and Schaer}

\begin{abstract}
This publication describes the motivation and generation of \texttt{$Q_{bias}$}, a large dataset of Google and Bing search queries, a scraping tool and dataset for biased news articles, as well as language models for the investigation of bias in online search.
Web search engines are a major factor and trusted source in information search, especially in the political domain.
However, biased information can influence opinion formation and lead to biased opinions.
To interact with search engines, users formulate search queries and interact with search query suggestions provided by the search engines.
A lack of datasets on search queries inhibits research on the subject.
We use \texttt{$Q_{bias}$} to evaluate different approaches to fine-tuning transformer-based language models with the goal of producing models capable of biasing text with left and right political stance.
Additionally to this work we provided datasets and language models for biasing texts that allow further research on bias in online information search.

\end{abstract}





\maketitle
\newpage
\section{Introduction}\label{sec:intro}
Search engines such as Google and Bing are seen as trustworthy sources of information on many topics, including political news and information~\cite{ray2020, edelman2022}.
Further, search engines have proven to have a major impact on the formation of political opinions~\cite{epstein2015}. 
To interact with search engines, users formulate search queries that are an expression of their information need.
Based on these queries search engines usually provide a set of search query suggestions~\cite{niu}.
Queries and the choices users make when interacting with query suggestions are based on the information need they want to satisfy, which is often to support their personal opinions or beliefs founded on previously encountered information~\cite{belkin2001}.
This interaction process and the results presented to the users are prone to be biased~\cite{Introna2000, Bolukbasi_2016, Mitra2014} and therefore the high level of trust can be seen as problematic.

However, the true effect of biased search queries and different types of inherent biases on the actual list of search results has not sufficiently been investigated. 
One of the main reasons for the infrequency of studies on bias in search queries might be a lack of publicly available datasets. 
datasets that include real-world user queries, query suggestions, and actual query reformulations are rare. 
One of the reasons is that collecting search queries from users is problematic due to privacy concerns.
Although there are techniques like pseudonymization, query logs enable the identification of users~\cite{Barbaro_aol}.
Previously available datasets such as the AOL query log dataset~\cite{aol} are no longer available, and their usage is morally debatable. 

Using unpersonalized search query suggestions as proxies might solve this issue.
Query suggestions describe the list of  predicted queries suggested to users during the input of search queries by the search engine, sometimes also called search predictions~\cite{sullivan_2018} or query auto completion~\cite{cai_survey_2016}.
While these search query suggestions can be generated locally from the result set \cite{xu_improving_2000} or be taken from global knowledge bases \cite{hienert_novel_2011}, in web search suggestions are mostly based on frequently issued queries by users~\cite{sullivan_2018}.
It can therefore be assumed that in many cases query suggestions are popular related search queries for the initial root query that represented the user's interest in a topic or entity.

The U.S. news domain is a popular domain for bias research due to its sociological relevance and the two-party left-right spectrum that facilitates bias analysis~\cite{robertson_auditing_2018}.
One of the benefits of our dataset is to enable in-depth investigations of the correlation between bias in search queries and search results, as search results for popular web search engines can easily be collected using our provided dataset of search queries.
Therefore, one goal of this work is to provide a large dataset of search query suggestions for the U.S. political news domain. Additionally, we provide a transformer-based methodology for biasing text.
Such a system can be a useful tool in bias research, f.e. for generating biased derivatives of search queries.
Thus, we evaluate a range of fine-tuning scenarios to find settings, that produce the most biased results.
We provide the fine-tuned transformer-based language models that are capable of inducing left and right political stance bias in the form of lexical biases.

This publication describes the motivation and generation of \texttt{$Q_{bias}$}, a large dataset of Google and Bing search queries, a scraping tool and dataset for biased news articles, as well as language models for the investigation of bias in online search. 

The main contributions of \texttt{$Q_{bias}$} and this paper are:
\begin{itemize} 
    \item Two datasets:\footnote{The datasets can be found at Zenodo (\url{https://doi.org/10.5281/zenodo.7682914})} (1) The (to the best of our knowledge) largest labeled dataset of search query suggestions of a single domain for Google and Bing. (2) A dataset of biased news articles of the U.S. political news domain. 
    \item A scraping tool that allows researchers to easily retrieve an up-to-date version of the biased news dataset.\footnote{\label{gith}\url{https://github.com/irgroup/Qbias}} 
    \item A new approach for producing biased search queries, using intentionally biased transformer-based language models capable of producing biased texts. 
    \item An evaluation of different fine-tuning settings to find the most capable setup for producing bias-inducing language models.
\end{itemize}

\section{Background and Related Work}\label{sec:relWork}
\paragraph{Definition of Bias is the News Domain}
Due to the subjective nature of bias, few publications attempt to explicitly define bias in news~\cite{D'Alessio_2000, chen-etal-2020-analyzing, Recasens_2013, allsides2}.
'Biased news' generally describes non-neutral and opinionated news, but both are fuzzy attributes~\cite{Recasens_2013}.
In the political domain, the term opinionated describes having a fixed political opinion and agenda, aspect of a cognitive (author- or reader-sided) bias often described as partisan bias~\cite{Gawronski_partisan} or (political) stance~\cite{liu2021}. 
Partisanship manifests at different granularities.
At publishing or reporting level, partisanship is expressed by the selection of certain topics called selection bias, and coverage bias, the selection of different views on a topic~\cite{D'Alessio_2000}.
At text level, statement bias describes ``members of the media interjecting their own opinions into the text''~\cite{D'Alessio_2000}, manifesting in overall opinionated texts. 
This can take various forms, ranging from phrasing bias, the use of non-neutral language~\cite{Hube_2018b} to moral framing and ideological bias~\cite{Mokhberian_2020}.
Spin bias describes bias introduced either by omitting necessary (neutral) information or adding unnecessary information~\cite{chen-etal-2020-analyzing}. 
At word and n-gram level, biases manifest as linguistic or lexical biases~\cite{chen-etal-2020-analyzing}.
Most linguistic biases are highly domain- and context-specific.
For example, framing bias describes subjective words, while epistemological bias can be attributed to words targeting the credibility of a statement~\cite{Recasens_2013}.
Since these biases can be identified more objectively, most approaches to detecting biases rely on the identification of linguistic biases~\cite{allsides2}.

\paragraph{Research on Bias in Online Search and Search Queries}
Few publications investigate bias in online search in aspects other than search results:
\citet{robertson_auditing_2018} investigate partisan bias and filter bubble effects in political searches by auditing SERPS for a set of queries.
They did not find significant evidence, that unbiased search queries in real search sessions performed by real users lead to filter bubble effects. 
The unanswered question is, whether biased queries in general lead to biased search results.

Few studies investigate bias in search queries.
Due to the difficult accessibility of biased search queries, most studies focus on bias in search query suggestions:
in most of those, the authors investigate topical group biases in search query suggestions in the political domain~\cite{Bonart_2019, haak_perception-aware_2021, HaakS22}.
Overall, they observe minor topical gender biases for search queries consisting of names of politicians.
Research on bias in online search has shown, that ``factors such as the topic of the query, the phrasing of query and the time at which a query is issued also impact the bias seen by the users''\cite{kulshrestha2019}.

\paragraph{Datasets of Biased News and Search Queries}\label{sec:datasets}
\citet{baly2020we} predict media bias using news articles collected from AllSides balanced news.\footnote{\url{https://www.allsides.com/unbiased-balanced-news}} 
AllSides news is a popular source for balanced news~\cite{baly2020we,chen-etal-2020-analyzing,Mokhberian_2020}, since AllSides has a high standard for assigning bias labels~\cite{allsides2}.
Similarly, \citet{chen-etal-2020-analyzing} use AllSides-labeled news articles and adfontes labels to analyze bias at different granularities.
\citet{Mokhberian_2020} develop a framing bias detection and quantification approach, using a collection of news articles that they label according to news outlet bias labels provided by AllSides.\footnote{\url{https://www.kaggle.com/datasets/snapcrack/all-the-news}}
datasets for search queries and query suggestions are sparse:
most are either not available anymore~\cite{aol} or focus on a narrow topic~\cite{Bonart_2019}.
\citet{robertson_auditing_2019} introduce recursive algorithm interrogation, a technique for recursively retrieving query suggestions of a root query and their consecutive suggestions.
This technique was employed by \citet{HaakS22} to investigate bias in the German political domain.
However, to the best of our knowledge, there currently is no large-scale dataset on search queries. 

\section{AllSides Scraper and Datasets}\label{sec:datasets}
\begin{figure}
  \includegraphics[width=\linewidth]{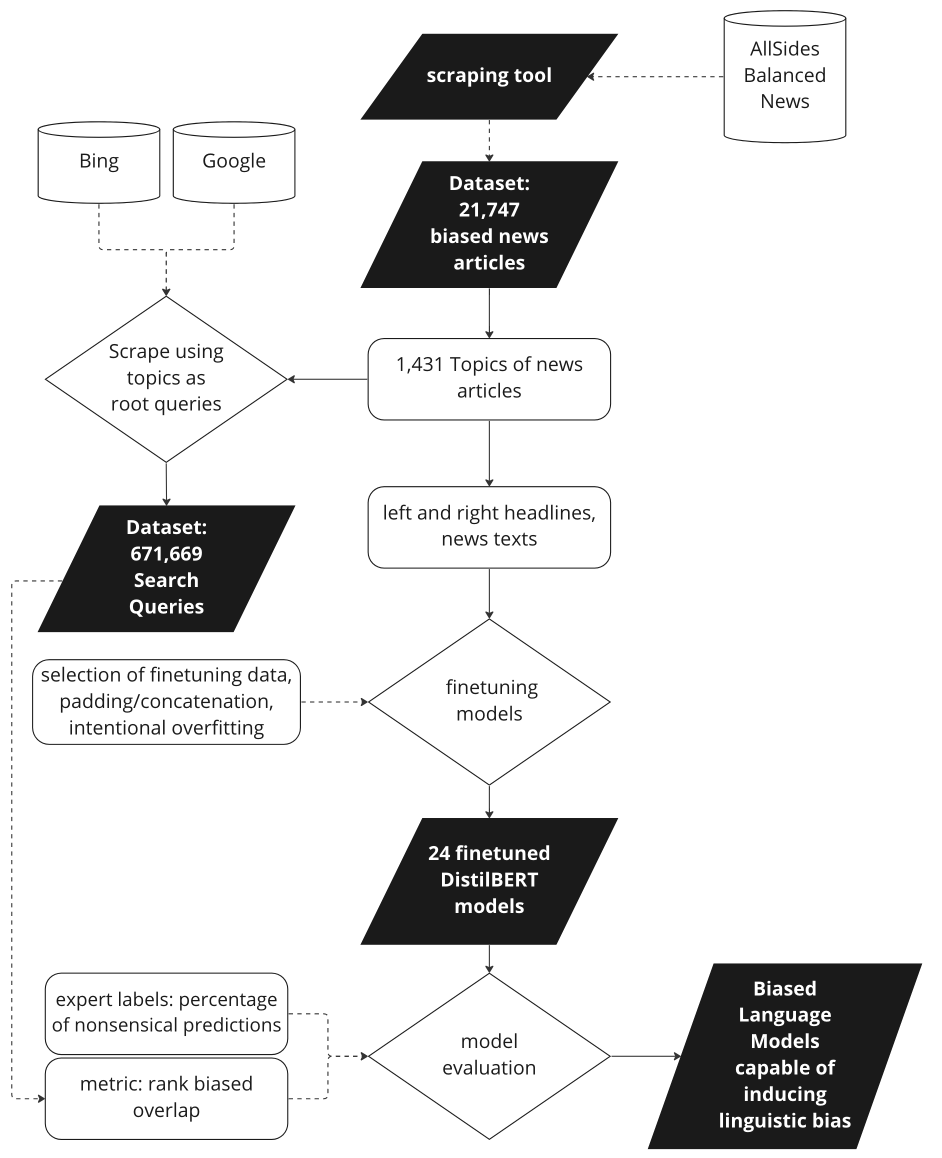}
  \caption{Pipeline used for generating bias-inducing language models. Black boxes represent datasets, language models (only available upon request), and tools provided via Zenodo, and GitHub.}
  \label{fig:flow}
\end{figure}

We present two novel datasets, that promote the investigation of bias in online news search. 
Both datasets can be found on Zenodo as mentioned in section~\ref{sec:intro}.
Further, we provide a web scraping tool for retrieving an up-to-date version of the AllSides dataset we provide. 
This section describes the content of the datasets as well as their creation process. 
Figure~\ref{fig:flow} shows how we assembled the datasets and use them in our approach for creating biasing language models.

\subsection{AllSides Scraper}\label{sec:crawler}
As described in section~\ref{sec:relWork}, the AllSides platform and the provides news is a frequently used source for high quality labeled biased news.
We want to provide an easy means of retrieving the news and all corresponding information.
Similar datasets have previously been produced, as mentioned in section \ref{sec:relWork}.
However, compared to the currently most recent available version by \citet{baly2020we}, our version includes more than 20 percent of additional articles.
Furthermore, for many tasks, especially in the news domain, it is relevant to have the most recent documents available.
We provide a Python-based scraper, that scrapes all available AllSides news articles and gathers available information as described in section~\ref{sec:allsides}.
By providing the scraper we facilitate access to a recent version of the dataset for other researchers.

\subsection{AllSides Biased News Dataset}\label{sec:allsides}
The dataset contains 21,747 news articles collected from AllSides balanced news headline roundups~\cite{allsides} in November 2022. 
The AllSides balanced news feature three expert-selected U.S. news articles from sources of different political views (left, right, center), often featuring spin bias, slant other forms of non-neutral reporting on political news~\cite{allsides2}. 
All articles are tagged with a bias label by four expert annotators based on the expressed political partisanship, left, right, or neutral~\cite{allsides2}.
The AllSides balanced news aims to offer multiple political perspectives on important news stories, educate users on biases, and provide multiple viewpoints~\cite{allsides2}. 
Collected data includes the headline, news text, publishing date, topic tags (e.g., "Republican party", "coronavirus", "federal jobs"), links to the article, and the publishing news outlet.
We also include AllSides' neutral description of the topic of the articles.  
Overall, the produced dataset contains 10,273 articled tagged as left, 7,222 articles tagged as right, and 4,252 articles tagged as center.
The collected articles have been published between June 2012, and the date of the data collection at the end of November 2022. 
We use the AllSides dataset for developing our biasing language model, as described in section~\ref{sec:LM}.
To allow for easier access to the most recent and complete version of the dataset for future research, we provide the scraping tool described in section~\ref{sec:crawler}.

\subsection{Search Queries in U.S. News Domain Dataset}\label{sec:queries} 
The second dataset we provide consists of 671,669 search query suggestions for root queries based on tags of the AllSides biased news dataset. 
We collected search query suggestions from Google and Bing for the 1,431 topic tags, that have been used for tagging AllSides news at least five times, approximately half of the total amount of topics. 
The topic tags include names, a wide range of political terms, agendas, and topics (e.g., "communism", "libertarian party", "same-sex marriage"), cultural and religious terms (e.g., "Ramadan", "pope Francis"), locations and other news-relevant terms.
On average, the dataset contains 469 search queries for each topic.
In total, 318,185 suggestions have been retrieved from Google and 353,484 from Bing.

Using a python implementation loosely adopting the framework provided by \citet{robertson_auditing_2019}, we scraped query suggestions for the topics as root queries.
Using Google Colab to run our scraper, we retrieved ten search query suggestions provided by the Google and Bing search autocomplete systems for the input of each of these root queries. 
Furthermore, we extended the root queries by the letters a to z (e.g., "democrats" (root term) $\rightarrow$ "democrats a" (query input) $\rightarrow$ "democrats and recession" (query suggestion)). 
The goal of this procedure is to simulate a user's input during information search.
Retrieving the suggestions for the root query and the extended queries generates a total of up to 270 query suggestions per topic and search engine.
The dataset we provide contains columns for root term, query input, and query suggestion for each suggested query.
The location from which the search is performed is the location of the Google servers running Colab, in our case Iowa in the United States of America, which is added to the dataset. 
Since we perform the scrape on a blank browser for each of the searches, personalization effects other than the location did not effect the suggested search queries.
Our scraping setup thus eliminates personalization effects that could, in theory, cause echo chamber effects~\cite{robertson_auditing_2018}.
Search engine providers describe that query suggestions are based on other users' searches~\cite{sullivan_2018}.
Successful attempts to influence query suggestions have confirmed this claim~\cite{wang2018}.
Thus, we deduce that query suggestions reflect real, frequently used search queries and can be used as proxies for search queries.
Despite the lack of information on the frequency of the collected search queries, our dataset contains the ranks of suggestions, as well as the suggestions to the root term for approximating the most frequent search queries.
Assuming that the topical tags in the AllSides news reflect popular topics, the produced dataset consists of real search queries for news-relevant topics.

\section{Developing Biasing Language Models}\label{sec:LM}
Another contribution \texttt{$Q_{bias}$} is to produce a pair of transformer-based language models that are capable of inducing left or right partisanship as linguistic biases.
This can be done by leveraging the masking function on the words of the target document. 
\subsection{Domain-Adopting DistilBERT}\label{sec:method_pipeline}
This section describes the methodological approach for developing transformer-based language models capable of biasing texts.
Usually, systems are developed with the goal of producing and reproducing as little bias as possible or to debias biased texts~\cite{raza_dbias_2022}.
We try to achieve the opposite, biasing texts by reproducing biases inherent in biased fine-tuning datasets.
This would, for example, allow for simulating a user that searches for biased terms potentially caused by exposure to biased news.
With the goal of capturing biased language in pretrained language models and using the HuggingFace Transformers module\cite{wolf-etal-2020-transformers}, we fine-tune the base DistilBERT model, which was trained in Wikipedia articles and a large book dataset.\footnote{\url{https://huggingface.co/distilbert-base-uncased}}
DistilBERT–\cite{distilBERT} was chosen as a base model since it has proven to perform well with comparably small datasets, performs better than BERT despite its smaller size~\cite{distilBERT}, and for its aptness for adopting domain-specificity~\cite{Bai2020ConstructionOD, Bykz2020AnalyzingEA}.
We are aware that there are more effective models, but to show the suitability of our approach, we chose DistilBERT due to its efficiency and sufficient effectiveness. 

We fine-tuned a range of pairs of models, each with one left model fine-tuned on the part of the AllSides corpus tagged as left-biased news and one right model fine-tuned with documents of the dataset labeled as right-biased news.
We produced 24 models, a left and a right model for 12 combinations of parameters.
Our goal is to find the ideal approach for capturing and reproducing as much bias as possible while producing meaningful results.
The central aspect we varied in fine-tuning the models is the data used.
The models are developed using three different data configurations: (a) the headline and news text of each of the left and right news articles of the AllSides dataset, (b) only the news text, and (c) only the headline.
Another factor we evaluate is the use of padding or concatenation to generate a consistent chunk size for fine-tuning. 
The chosen chunk size is the max length of fine-tuning documents for the padding approach and 128 for the concatenation approach.
Lastly, we compare the effects of intentional overfitting by raising the number of epochs to 20, compared to a more reasonable 6 epochs, which proved to produce a good combination of training loss and validation loss for all dataset configurations.
In theory, overfitting could be another possibility for reproducing biased formulations from the texts used in fine-tuning. This is why we tried to intentionally raise the number of epochs as a parameter~\cite{mosbach2021on}. 

\subsection{Evaluating the Biasing Language Models}
All 24 models are available upon request.
We decided not to make the models publicly available due to concerns about potential harms that could be caused by misuse of the systems that we further elaborated in section \ref{sec:moral}.
For evaluating the models' effectiveness, we choose a combination of manual assessment and quantitative measures.
The task of the models is to generate biased output, however, biases are diverse and not easily measurable.
Since we have no way of identifying biased tokens in the dataset, perplexity is not a useful measure for evaluating the model. 
Further, we cannot effectively measure bias as a criterion for the effectiveness of the ability of the systems to produce biased versions of queries.
However, we can assume that when masking words in search queries for topics of the political news domain and letting the pairs of language models predict the words, the output of the left and right models should differ from each other.
We choose to evaluate their performance on search queries since we want to use the models in future research on bias in search results for biased and unbiased search queries.
Further, they represent texts of a different type but the same domain as the texts used in training the models.

For each pair of models, we measure the difference between the two models' 10 most probable non-punctuation predictions for 100 randomly selected queries from the dataset by measuring the Rank Biased Overlap (RBO) with $p = 0.9$.
We mask a random token of each query that is neither part of the original news topic nor a stopword since we want to let the models generate meaning-carrying tokens while keeping the query's topic intact.
Although nonsensical random predictions or neologisms would lead to great RBO scores, we want to generate queries, that humans could have formulated. 
To assure that the models generate output that makes sense, we let the models generate next word predictions for ten bias-provoking sentences (e.g., "Hilary Clinton is a", "Covid Vaccines should be").
Two domain experts (a professor and a postdoctoral researcher) then label nonsensical and biased predictions. We measure the inter-annotator agreement on the individual statements (independent of the models, that produce the statements) with Cohen's Kappa~\cite{cohensKappa}. For nonsensical predictions, we obtained a value of 0.18 (slight agreement), for biased predictions the value is 0.84 (almost perfect).
The bias labels are assigned if the suggested word induces political stance bias.

Table~\ref{tab:evaluation} shows the results of the model evaluation process. 
The lower $h$,$t$, or $ht$ describes if headline, text, or both have been used.
$RBO$ is the rank bias overlap between the left and the right model of each configuration.
$P_{nons}$ describes the average percentage of nonsensical predictions of both models and $P_{bias}$ the percentage of biased predictions. 
Overall, the models fine-tuned with only headlines show the best (lowest) RBO scores, as well as the lowest $P_{nons}$ scores.
Headlines and text and only text perform more or less equally in terms of their RBO. 
Concatenating texts produces on average better results than padding, although for fine-tuning with only headlines, the effect is minimal. 
Intentional overfitting by raising the amount of training epochs seems to worsen the RBO scores.
The percentage of nonsensical predictions does not differ much between different fine-tuning setups, fine-tuning on texts only seems to be the only scenario that increases the percentage of nonsensical predictions.
Many of the models generate a high percentage of biased suggestions, with the headline models having the highest percentage of biased predictions. 

\begin{table}[t]
\caption{Evaluation results of the fine-tuned language models. The highlighted values are the best results for each metric.}
  \label{tab:evaluation}
\begin{tabular}{l|c|c|c}
Model configuration & $RBO$  & $P_{nons}$ & $P_{bias}$\\ \hline
$DistilBERTconcat_{h}$ & \textbf{0.54} & 0.04 & 0.33 \\\hline
 $DistilBERTconcat_{t}$&  0.72& 0.10 & 0.14 \\\hline
$DistilBERTconcat_{ht}$ & 0.68 & 0.08 & 0.15 \\\hline
$DistilBERTconcat_{h}+overfitting$ & 0.56  & \textbf{0.03} & \textbf{0.5} \\\hline
$DistilBERTconcat_{t}+overfitting$ & 0.73 & 0.11 &  0.14 \\\hline
$DistilBERTconcat_{ht}+overfitting$ & 0.68 & 0.07  & 0.23 \\\hline
$DistilBERTpadding_{h}$& \textbf{0.54} & 0.04  & 0.38\\\hline
$DistilBERTpadding_{t}$& 0.74 &  0.05 & 0.28 \\\hline
$DistilBERTpadding_{ht}$& 0.77 &  0.07 & 0.1 \\\hline
$DistilBERTpadding_{h}+overfitting$ & 0.58 & \textbf{0.03} & 0.43 \\\hline
$DistilBERTpadding_{t}+overfitting$& 0.77 & 0.06 & 0.24 \\\hline
$DistilBERTpadding_{ht}+overfitting$ &0.77  & 0.07 & 0.12 \\\hline

\end{tabular}
\end{table}

\section{Discussion}

Our language models show, that by using small, high-quality datasets, it is possible to fine-tune transformer-based language models to bias texts.
In our fine-tuning setup, models fine-tuned with headlines produce the overall best results. 
As the main reason for that, we assume that the high amount of quotes in news texts, which often are statements the authors of the articles disagree with, might have induced noise in fine-tuning the models. 
Further, the condensed nature of headlines, which aims to catch attention, might also reflect in resulting language models.
The output of left and right models fine-tuned with headlines produce texts containing linguistic biases.
For example, for "Donald Trump is a", our model produces "hero" as right-biased and "fraud" as left-biased next word predictions.
The overall low percentage of nonsensical predictions supports RBO as a suited evaluation metric.
Despite the overall good results, future research should investigate the findings with other language models and compare the performance to our results.

Our scraping tool and the datasets not only allow us to effectively generate the language models but enables the investigation of other research topics in the political news information search domain such as bias classification or sentiment analysis.

\paragraph{Remark on Moral Issues}\label{sec:moral}
We are aware, that building bias-inducing systems and providing datasets that enable the reproduction of developing similar systems is problematic.
We also did consider that our work can inspire and help to develop transformer-based language models capable of producing biased, toxic, or otherwise harmful texts.
The severity of openly published biased systems has been shown by other such models, e.g. the gpt-4chan model.\footnote{\url{https://huggingface.co/ykilcher/gpt-4chan}}
The publication of the model incited a debate, that shows how problematic biased language made available to a wide public audience are, despite explicitly stating the intended use for research applications~\cite{4chan_reddid,4chan_vice}.

Our models are biased primarily in terms of linguistic biases that reflect political stances and views on political topics.
Since this can in part include objectively wrong and opinionated statements, hate speech, racism, and other forms of despicable language and toxicity, our models can and will reproduce text, that conveys these phenomena. 
To minimize the harmful effects of our publication that could be caused by the misuse of the models, we decided to not make our biased models publicly accessible.
Despite that, we provide access to the models upon request for research, if a reasonable application use case is provided.
However, we do so in the interest of investigating biases and correlations of bias in online information search, with the goal of increasing transparency and fairness.
Raising awareness for how easily, intentionally or not, biases can be reproduced and induced with AI systems is part of this endeavor.

Although we have shown how the data we provide can be used to create models that produce biased language we chose to publish the datasets and scraper.
This is mainly because of three reasons: (a) as stated in section \ref{sec:datasets} similar datasets are available publicly, (b) we believe that the benefits of raising awareness outweigh making already public biased data more accessible, and (c) datasets on biased language are required for developing systems to detect and inhibit bias. 
With our work, we hope to highlight the need to assure, that data used for developing models is unbiased and raise awareness for how easily transformer-based language models can be fine-tuned to produce biased language.

\section{Outlook}
This publication represents the first and foundational milestone of a larger-scale investigation of bias in online information search. 
As a major next contribution, we plan to use the presented datasets and models to investigate the effects of biased and unbiased search queries on the search results of different search engines for popular topics of the U.S. political news domain.
To accomplish this, we need to overcome the issue of subjectivity and lack of effective methodological approaches to bias identification other than by employing human annotation.
We plan to introduce an ensemble of methods, including bias-agnostic analysis of linguistic differences, lexical features, and transformer-based approaches for bias classification.
Further, we plan to conduct a study using a simulation approach.
By simulation different user behaviors in terms of information need, formulating queries, and interacting with query suggestions in an interactive information search simulation, we plan to gain insights into echo chamber effects and bias formation in information search.
Additionally, we hope to identify which user properties and behaviors increase and which lower the risk of encountering bias in information search. 

\section{Conclusion}
This work presents \texttt{$Q_{bias}$}, a first milestone of our ongoing research on bias in online search.
We present two datasets: a biased news dataset and a large dataset of biased and unbiased search queries for topics of the U.S. political news domain. 
Further, we provide a scraping tool, that allows for collecting bias-labeled news texts from AllSides. 
Lastly, we evaluate approaches to fine-tuning DistilBERT transformer-based language models for biasing texts and publish our models capable of inducing left and right political stance bias in the form of lexical biases.

\bibliographystyle{ACM-Reference-Format}
\bibliography{main}


\end{document}